\documentclass[12pt]{article}

\newcommand{\x}{arXiv:}
\newcommand{\m}{\mathrm}

\usepackage{graphicx}

\begin{document}
\thispagestyle{empty}
\begin{center}

\null \vskip-1truecm \vskip2truecm

{\Large{\bf \textsf{Shearing Black Holes and Scans of the Quark Matter Phase Diagram}}}

{\Large{\bf \textsf{}}}

{\large{\bf \textsf{}}}

{\large{\bf \textsf{}}}

\vskip1truecm

{\large \textsf{Brett McInnes
}}

\vskip1truecm

\textsf{\\ NORDITA,  KTH Royal Institute of Technology and Stockholm University}
\vskip0.1truecm
\textsf{\\ Roslagstullsbacken 23, SE-106 91 Stockholm, Sweden}
\vskip0.1truecm
\textsf{\\ and}
\vskip0.1truecm
\textsf{\\ National
  University of Singapore}\footnote{Permanent Address}
  \vskip0.3truecm
\textsf{email: matmcinn@nus.edu.sg}\\

\end{center}
\vskip1truecm \centerline{\textsf{ABSTRACT}} \baselineskip=15pt

\medskip
Future facilities such as FAIR and NICA are expected to produce collisions of heavy ions generating quark-gluon plasmas with large values of the quark chemical potential; peripheral collisions in such experiments will also lead to large values of the angular momentum density, associated with the internal shearing motion of the plasma. It is well known that shearing motions in fluids can lead to instabilities which cause a transition from laminar to turbulent flow, and such instabilities in the QGP have recently attracted some attention. We set up a holographic model of this situation by constructing a gravitational dual system exhibiting an instability which is indeed triggered by shearing angular momentum in the bulk. We show that holography indicates that the transition to an unstable fluid happens more quickly as one scans across the quark matter phase diagram towards large values of the chemical potential. This may have negative consequences for the observability of quark polarization effects.

\newpage
\addtocounter{section}{1}
\section* {\large{\textsf{1. Quark Polarization and AdS Black Holes}}}
The quark matter phase diagram \cite{kn:behan}\cite{kn:ohnishi}\cite{kn:enrod}, representing states of quark matter at various values of the temperature and quark chemical potential, is currently understood only in an approximate or qualitative sense. Until recently, experimental investigations, based on the study of heavy ion collisions, have only explored the region of the diagram near to the temperature axis; but a move towards larger values of the quark chemical potential is in its early stages in the beam energy scan programme of the RHIC collider \cite{kn:ilya}\cite{kn:dong}, and will reach maturity with the completion of the FAIR \cite{kn:fair} and NICA \cite{kn:nica} facilities.

\begin{figure}[!h]
\centering
\includegraphics[width=1.25\textwidth]{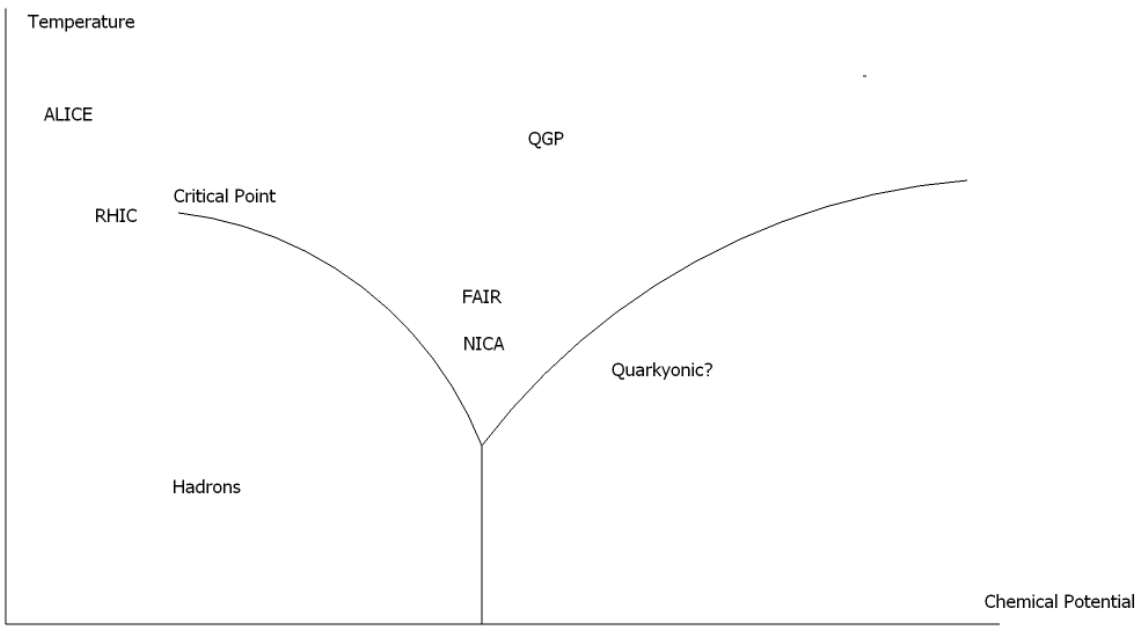}
\caption{Conjectured Quark Matter Phase Diagram}
\end{figure}

In broad terms, one version\footnote{For other possible versions, see \cite{kn:fuku}.} of the phase diagram has the following structure. As in more familiar examples of phase diagrams, there are several lines separating various phases, with the Quark-Gluon Plasma [QGP] occupying all regions at sufficiently high temperature, and hadronic matter being found at lower temperatures and low values of the chemical potential. There is thought to be a critical point at a temperature T$_{\m{c}}$ given approximately by T$_{\m{c}}$ $\approx$ 175 MeV and at a value of the quark chemical potential $\mu$ which is unknown but estimated to be in the range 1 - 4 times T$_{\m{c}}$. The phase line separating the QGP and hadronic phases is thought to bend \emph{downwards} from this point  \cite{kn:katz}\cite{kn:enrodi}, so that the QGP can exist, at relatively high values of $\mu$, at substantially \emph{lower} temperatures than would be possible at or near the $\mu$ = 0 axis. The phase line continues downward until it reaches a triple point, at a location which is unknown; the nature of the third phase [which may in fact consist of several distinct phases] is likewise unknown [see for example \cite{kn:andronic}, and \cite{kn:borun} for a holographic approach].  FAIR and NICA are designed to explore the region of the phase diagram around and beyond the critical point, and perhaps ultimately towards the triple point. Naturally it is of great interest to have theoretical predictions as to what will happen as the chemical potential is increased in this manner. We now turn to a particular phenomenon which might be affected.

When heavy ions collide peripherally, a very simple and compelling kinematic argument shows that, for favourable impact parameters, large quantities of angular momentum \emph{must} be transferred to the QGP, should it be formed \cite{kn:liang}\cite{kn:bec}\cite{kn:huang}. This is based on the simple fact that the nucleon density in each ion is non-uniform in the transverse directions, resulting in a \emph{shearing} motion within the QGP formed in a peripheral collision. [See \cite{kn:bec} for a particularly clear discussion.]  This can be expected to have important indirect consequences [for studies of the vorticity of the plasma \cite{kn:bemo}\cite{kn:leigh}\cite{kn:sorin}], but it might also have directly observable implications. It has been very plausibly argued \cite{kn:liang}\cite{kn:bec}\cite{kn:huang} that spin-orbital coupling  will under these circumstances entail global quark polarization effects, leading to several striking and characteristic observable phenomena.

Thus far, no unambiguous signals of this kind have been reported by the RHIC experiments: see for example \cite{kn:STAR}\cite{kn:RHIC} for searches for such effects, specifically for spin alignment of certain kinds of vector mesons. Unfortunately it is very difficult \cite{kn:jianhua} to estimate the precise value of the angular momentum at which observations might be expected, and there may be additional effects suppressing the polarization \cite{kn:ayala}. However, Becattini et al. \cite{kn:bec} estimate that the QGP formed in collisions of gold nuclei at the RHIC might, for an optimal impact parameter, acquire a very large angular momentum of 7.2 $\times$ 10$^4$ per collision in natural units, and the corresponding figure at the LHC could be significantly larger. Thus, the failure to observe polarization is somewhat puzzling.

In these discussions, however, it is implicitly assumed that the shearing flow of the plasma is laminar. This is a somewhat dubious assumption, since one of the most basic phenomena in hydrodynamics is precisely the transition from laminar to turbulent flow triggered by shear \cite{kn:drazin}\cite{kn:lin}. Classically, for example, one expects \emph{Kelvin-Helmholtz instability} to evolve, in many situations, when fluids are sheared, and indeed the question as to whether this instability arises in the shearing flow of the QGP has recently been investigated \cite{kn:KelvinHelm}. It is also entirely possible that the QGP develops other, non-classical hydrodynamic instabilities. Clearly, if that is so, and \emph{if the transition to turbulence occurs quickly}, there could be serious consequences for polarization effects, which take some time to develop \cite{kn:huang}.

In this work, we try to determine whether holographic methods \cite{kn:solana}\cite{kn:pedraza}\cite{kn:youngman} can throw any light on this situation. We argue that this may well be the case, as follows.

As always, one begins by choosing a thermal AdS black hole spacetime, which is to be dual to a field theory representing the plasma. Clearly we need a bulk black hole endowed with angular momentum. One first thinks of the AdS-Kerr metric  \cite{kn:carter}\cite{kn:hawrot}, and this metric has indeed been used to good effect by Atmaja and Schalm \cite{kn:schalm} to study certain aspects of a QGP endowed with large quantities of angular momentum. However, the conformal boundary in this case does not rotate differentially; the angular momentum is due to rotation, not shear. In order to obtain a more realistic geometry for the dual theory one needs to use the ``rotating'' \emph{planar} black holes discovered by  Klemm, Moretti, and Vanzo \cite{kn:klemm} [the ``KMV$_0$ black holes''\footnote{The subscript indicates planar topology for the event horizon, and planar spatial geometry at infinity.}]. As we shall explain below, these black holes do not actually ``rotate'': instead, there is a shearing motion within the sections transverse to the radial direction, including the one at infinity. This describes a shearing, \emph{laminar} flow of the QGP; which is precisely what we need.

Next, we need to find a mechanism which causes the bulk to become unstable: this should have a well-understood holographic interpretation and, most importantly, \emph{it must be possible to show that this instability is triggered by the shearing motion at infinity} [but \emph{not} by rotation there, which would not induce Kelvin-Helmholtz instability]. We do this by borrowing a mechanism from string theory. We place a certain configuration of BPS branes in the bulk, setting up a system studied in detail by Seiberg and Witten \cite{kn:seiberg}. This system is stable in the absence of shearing [even if there is rotational motion at infinity] but becomes unstable \cite{kn:75}\cite{kn:76} when shearing is present: the brane action becomes unbounded below. Thus, this mechanism, which has a well-understood dual, is a good candidate for a holographic account of the Kelvin-Helmholtz instability in the QGP.

A basic observation in all studies of the QGP, as it is produced in heavy-ion collisions, is that its lifetime is very short ---$\,$ it is estimated that local thermal equilibrium is established around 1 fm/c, and that the QGP hadronizes at around 5 fm/c after the collision \cite{kn:ipp} at the RHIC, though somewhat later at the LHC ALICE experiment \cite{kn:reygers}. The most important question in studies of possible QGP instabilities is therefore: how long does it take for the instability to develop? In view of the fact that a true dual description of QCD is not known \cite{kn:mateos}\cite{kn:karch}\cite{kn:kovch}, we feel that holographic techniques are not yet in a position to answer such questions. Secondary, more qualitative questions, however, can be addressed: does the time taken depend on other variables? If so, in what way?

Our main result is that our holographic model of shearing instabilities in the QGP indicates that the time required \emph{does} depend quite strongly on the quark chemical potential\footnote{For a phenomenological study of the role of the chemical potential in connection with quark polarization, see \cite{kn:shipu}.}, represented holographically in the usual way \cite{kn:chamblin}\cite{kn:koba} by adding electric charge to the black hole. In particular, the instability arises much more quickly at very high values of that parameter. In short, the transition to turbulence may be more marked in the future experiments discussed above, at FAIR and NICA, with possible negative consequences for the observability of quark polarization effects in those experiments.

We begin with a discussion of the relevant black hole spacetime.

\addtocounter{section}{1}
\section* {\large{\textsf{2. The QKMV$_0$ Spacetime}}}
As is well known, in the asymptotically AdS case it is possible to find black hole spacetimes with event horizons having non-spherical topology \cite{kn:lemmo}. This is fortunate, since it opens the way to considering boundary field theories defined on topologically trivial spacetimes, such as those in which actual heavy ion collisions occur. Evidently we need black holes with topologically \emph{planar} event horizons, but we begin with non-rotating charged [AdS-Reissner-Nordstr$\ddot{\m{o}}$m] black holes of arbitrary event horizon topology.

If the event horizon is an n-dimensional [n $\geq$ 2]  compact space of constant curvature k = $\{ -1, 0, +1 \}$ [so that, in the k = 0 case, we are dealing with a torus or related manifold], then the (n + 2)-dimensional metric with asymptotic AdS curvature  $-1$/L$^2$ is given by
\begin{eqnarray}\label{ALEPH}
g(AdSRN^{\m{k}}_{n+2}) &=& -\, \Bigg[{r^2\over L^2}\;+\;\m{k}\;-\;{16\pi M\over n V[X^{\m{k}}_n] r^{n-1}}\;+\;{8\pi Q^2\over n(n-1)\big(V[X^{\m{k}}_n] \big)^2 r^{2n-2}}\Bigg]dt^2\;
\nonumber \\
& & + \;{dr^2\over {r^2\over L^2}\;+\;\m{k}\;-\;{16\pi M\over n V[X^{\m{k}}_n] r^{n-1}}\;+\;{8\pi Q^2\over n(n-1)\big(V[X^{\m{k}}_n] \big)^2 r^{2n-2}}} \;+\; r^2\,d\Omega^2[X^{\m{k}}_n],
\end{eqnarray}where $d\Omega^2[X^{\m{k}}_n]$ is a metric of constant curvature k = $\{ -1, 0, +1 \}$ on an n-dimensional space $X^{\m{k}}_n$ with [dimensionless] volume $V[X^{\m{k}}_n]$, and $M$ and $Q$ are the ADM mass and charge respectively \cite{kn:peldan}. The densities of $M$ and $Q$ at the event horizon of the hole are given, if $r_h$ is the value of the radial coordinate r at the event horizon, respectively by $M/V[X^{\m{k}}_n]r^n_h$ and $Q/V[X^{\m{k}}_n]r^n_h$.

If we define $M^* \;=\;M/V[X^{\m{k}}_n],\;Q^*\;=\;Q/V[X^{\m{k}}_n]$, then $r_h$ is determined by $M^*$ and $Q^*$, and the same is true of the densities, which are given by $M^*/r^n_h$ and  $Q^*/r^n_h$. In the k = 0 case we can now replace the torus by a space of \emph{planar} topology: let the volume factor tend to infinity, along with $M$ and $Q$, in such a way that the quotients $M^*$ and $Q^*$ remain finite\footnote{Note that it is $Q^*$ that appears in the formula for the corresponding electromagnetic potential one-form.}. Thus we deduce metrics for black holes with event horizons having planar topology. [This argument is inspired by Witten's derivation \cite{kn:confined} of a planar black hole metric from a spherical one.]

In short, four-dimensional Planar AdS-Reissner-Nordstr$\ddot{\m{o}}$m black holes have  metrics of the form
\begin{equation}\label{ALPHA}
g(PAdSRN) = -\, \Bigg[{r^2\over L^2}\;-\;{8\pi M^*\over r}+{4\pi Q^{*2}\over r^2}\Bigg]dt^2\; + \;{dr^2\over {r^2\over L^2}\;-\;{8\pi M^*\over r}+{4\pi Q^{*2}\over r^2}} \;+\;r^2\Big[d\psi^2\;+\;d\zeta^2\Big],
\end{equation}
where $\psi$ and $\zeta$ are dimensionless coordinates on the plane. Let us now add angular momentum. It is described by a parameter, conventionally denoted $a$, defined as the ratio of the angular momentum density of the black hole [defined now by a suitable limit] to its mass density. The corresponding black hole metrics, which we have called the ``KMV$_0$ metrics'', were obtained by Klemm, Moretti, and Vanzo \cite{kn:klemm}; with the addition of electric charge, which is straightforward, we may call these the ``QKMV$_0$ metrics'':
\begin{equation}\label{BETA}
g(QKMV_0) = - {\Delta_r\Delta_{\psi}\rho^2\over \Sigma^2}\,dt^2\;+\;{\rho^2 \over \Delta_r}dr^2\;+\;{\rho^2 \over \Delta_{\psi}}d\psi^2 \;+\;{\Sigma^2 \over \rho^2}\Bigg[\omega\,dt \; - \;d\zeta\Bigg]^2,
\end{equation}
where the asymptotic curvature is $-1$/$L^2$, and where
\begin{eqnarray}\label{GAMMA}
\rho^2& = & r^2\;+\;a^2\psi^2 \nonumber\\
\Delta_r & = & a^2+ {r^4\over L^2} - 8\pi M^* r + 4\pi Q^{*2}\nonumber\\
\Delta_{\psi}& = & 1 +{a^2 \psi^4\over L^2}\nonumber\\
\Sigma^2 & = & r^4\Delta_{\psi} - a^2\psi^4\Delta_r\nonumber\\
\omega & = & {\Delta_r\psi^2\,+\,r^2\Delta_{\psi}\over \Sigma^2}\,a.
\end{eqnarray}
Notice that (\ref{BETA}) reduces to (\ref{ALPHA}) when $a$ = 0, so this is the generalization of the Planar AdS-Reissner-Nordstr$\ddot{\m{o}}$m geometry to allow for the presence of angular momentum. Notice too that the metric induced on $t$ = constant sections at the horizon has determinant equal to $r^4_h$; therefore, $M^*$ and $Q^*$ retain their interpretations as in (\ref{ALPHA}).

With obvious changes, the properties of this geometry on and outside the horizon closely resemble those of the zero-charge  KMV$_0$ spacetime, and the reader is referred to \cite{kn:klemm} for the details. Here we discuss only those properties directly relevant to constructing a holographic model of a dual field theory with a large angular momentum density and chemical potential.

The QKMV$_0$ geometry induces at infinity a conformal structure represented by a metric of the ``peripheral collision'' \cite{kn:75} form:
\begin{equation}\label{DELTA}
g_{\rm PC} \;=\; - \, \m{d}t^2 \;-\; 2\omega_{\infty}(x)\, L \, \m{d}t\m{d}z \;+\; \m{d}x^2 \;+\; \m{d}z^2.
\end{equation}
Here $x$ and $z$ are Cartesian coordinates\footnote{We are following the standard conventions of the heavy-ion literature [see for example \cite{kn:bec}]: the $z$ axis is the axis of the collision, the $x$ axis is transverse, and the system is analysed sectionally, that is, each $y$ = constant plane is considered separately, so that the system is effectively two-dimensional.}; they are related to the coordinates $\psi$ and $\zeta$ in equation (\ref{BETA}) by d$x$ = $L$d$\psi$/$\sqrt{1 + a^2 \psi^4/L^2}$, d$z$ = $L$d$\zeta$. [We can take $x \geq 0$ and $z \geq 0$ in this case.] The function $\omega_{\infty}(x)$ is the asymptotic value of the angular velocity of the black hole. Free particles in such a spacetime, with $x$ = constant and zero momentum in the $z$ direction, move in the positive $z$ direction [as a result of frame-dragging] at a dimensionless speed given by
\begin{equation}\label{EPSILON}
v(x) \; \equiv \; \m{d}z/\m{d}t = \omega_{\infty}(x)L.
\end{equation}
For example, in the specific case of the QKMV$_0$ geometry, $v(x)$ is given by
\begin{equation}\label{ZETA}
v(x) \;=\; a\psi^2/L.
\end{equation}
This is indeed a non-trivial function of $\psi$, and therefore of $x$, so there is a \emph{shearing}, not a rotational motion in this case\footnote{Notice that this equation implies that $a > 0$, since $v(x)$ is certainly positive away from the axis.}. That is precisely what is needed to describe the internal motion of the QGP early in the hydrodynamic regime in the aftermath of a peripheral heavy-ion collision.

It is important to note that while the full boundary metric is not flat, it is [globally] \emph{conformally} flat: this was verified explicitly in \cite{kn:76}. Thus, as far as conformally coupled fields are concerned, the boundary geometry is effectively Minkowskian; any misbehaviour on the part of such fields cannot be ascribed to the unusual geometry [as can happen when the boundary is not conformally flat \cite{kn:seiberg}]. This is closely similar to the situation described in \cite{kn:sonner}, where a rotating superconductor is described holographically using the boundary of the topologically spherical AdS-Kerr-Newman black hole: this boundary too has a non-trivial spacetime geometry, but it is [locally] conformally flat.

We stress that this result would \emph{not} have been obtained if we had used the topologically spherical AdS-Kerr-Newman geometry. The latter does have differential rotation rates outside the event horizon, \emph{but this effect fades away} at large distances, and the topological sphere at infinity rotates [like the event horizon] as if it were a rigid body \cite{kn:cognola}.

To summarize: the transverse sections [r = constant] of the QKMV$_0$ black hole are not really ``rotating" in the conventional sense: they are \emph{shearing}, in the sense that the frame-dragging effect mimics a shear, and ---$\,$ crucially ---$\,$ this effect persists \emph{even at infinity}. The QKMV$_0$ black hole is therefore well suited to provide a holographic description of the QGP when it is endowed with non-trivial quantities of angular momentum; which is in fact the generic case.

Now let us turn to the question of relating the QKMV$_0$ parameters to those of the QGP when the latter's angular momentum density and quark chemical potential are both non-zero.

\addtocounter{section}{1}
\section* {\large{\textsf{3. Holography of the QKMV$_0$ Parameters }}}
The value of $r$ at the event horizon of the QKMV$_0$ black hole, $r_h$, is obtained by setting $\Delta_r$ above equal to zero and taking the largest positive root:
\begin{equation}\label{EPSILON2}
a^2\;+\;4\pi Q^{*2} \; - \; 8\pi M^*r_{h} \; + \;r_{h}^4/L^2 \; = \; 0.
\end{equation}

We see that, as far as the position of the event horizon is concerned, the effect of adding charge is just to augment the effect of angular momentum. Quantities associated with the event horizon, such as the Hawking temperature, therefore depend on charge and angular momentum in much the same way.

As usual, the root only exists if the constant term, given by a combination of the angular momentum and charge parameters, is not too large: we have
\begin{equation}\label{eq:EPSILONEPSILON2}
{a^2\;+\;4\pi Q^{*2} \over L^2} \; \leq \; 3\times (2\pi M^*/L)^{4/3}.
\end{equation}
Notice that this is the only restriction on the parameter $a$. [This is in sharp contrast to the topologically spherical case, where, independently of the censorship condition, $a$ is bounded by the curvature scale $L$.]

The temperature of the black hole is given by a complicated combination of $M^*$, $Q^*$, and $a$, but for our purposes it will be more useful to eliminate $Q^*$ and $a$, so as to express the temperature in terms of  $M^*$ and $r_h$ only [as in \cite{kn:klemm}]:
\begin{equation}\label{ZETA2}
T\;=\;{r_h\over\pi L^2}\;-\; {2M^* \over r^2_h}.
\end{equation}
One can readily verify that this decreases monotonically as either charge or angular momentum is added to a black hole with fixed $M^*$, and that it vanishes for all extremal holes.

Let us assume that the QKMV$_0$ black hole has a dual interpretation, in the usual holographic sense. We now ask whether it is possible to reconstruct the black hole geometry, given the relevant physical parameters on the boundary.

If $Q^*$ = 0, then, if we know $a, T,$ and $L$, we can completely reconstruct the geometric parameters of the dual black hole by solving the two equations (\ref{EPSILON2}) and (\ref{ZETA2}) for the two unknowns $r_h$ and $M^*$. But if $Q^*$ does not vanish, then we need a third equation. This is provided by the relation between the charge parameter and the chemical potential $\mu$. [Note that we are referring here to the quark, not the baryonic, chemical potential.]

There is a standard procedure for establishing the relation between the charge parameter and $\mu$, but here it encounters a peculiar difficulty. Recall that, by definition, a stationary black hole spacetime has an asymptotically timelike Killing vector $\partial_t$. Usually this vector field vanishes at certain points on the event horizon: the Killing flow can have fixed points there. Since the zero vector is null, this can in fact only happen where the ergosphere [defined as the locus where $\partial_t$ becomes null] touches the event horizon, as happens, for example, at the poles of the topologically spherical AdS-Kerr-Newman black hole. Now if the electromagnetic potential vector $A$ is to be regular, it follows that its time component, $A_t$ = $A(\partial_t$), must vanish at those points, if they exist. This determines the leading term in the asymptotic expansion of $A$, and that fixes $\mu$.

The problem here is that, for the QKMV$_0$ spacetime, the event horizon \emph{never} intersects the ergosurface \cite{kn:klemm}.
The proof of this statement is straightforward: with some algebraic manipulations one can write the [Lorentzian] norm of $\partial_t$ as
\begin{equation}\label{ETA2}
 g(QKMV_0)(\partial_t,\partial_t)\;=\;{- \Delta_r \;+\;a^2\Delta_{\psi} \over \rho^2},
\end{equation}
which clearly never vanishes on the event horizon if $a$ is not zero, since $\Delta_{\psi}$ is never zero. This means that $\partial_t$ never vanishes anywhere on the event horizon, and one can readily show that the same is true of the other Killing vector, $\partial_{\zeta}$. Thus the electromagnetic potential is completely regular everywhere on and outside the event horizon, provided that its components are so. The potential vector in this case is given by\footnote{The KMV$_0$ metric was obtained in \cite{kn:klemm} as a special case of the Pleba\'nski--Demia\'nski family of metrics [see \cite{kn:exact}]. The form of the electromagnetic potential for charged Pleba\'nski--Demia\'nski metrics is known, and the form in (\ref{LAMBO}) is obtained from it by setting the ``acceleration'' and NUT parameters equal to zero.}
\begin{equation}\label{LAMBO}
A = \Bigg[-{Q^*r \over  \rho^2}+\kappa_t\Bigg]\m{d} t \;-\; \Bigg[{aQ^*r\psi^2 \over  \rho^2}+\kappa_{\zeta}\Bigg]\m{d}\zeta,
\end{equation}
where $\kappa_t$ and $\kappa_{\zeta}$ are arbitrary constants. From the definition of $\rho$ given above, one sees that the coefficients here are regular everywhere on and outside the event horizon, and so the potential vector for these black holes is automatically completely regular, for any values of $\kappa_t$ and $\kappa_{\zeta}$. Thus the requirement of regularity yields no information.

We can solve this problem by requiring regularity not just of $A$, but also of its Euclidean version, $A^E$. [This is normally given as an alternative derivation of the relation between the charge and $\mu$; it is a natural approach, in view of the fact that the regularity of the Euclidean \emph{metric} fixes the relationship between the temperature and the mass of the black hole.] We begin by examining equation (\ref{ZETA}), the fundamental relation between frame-dragging and the velocity profile of the shearing fluid.

Note first that equation (\ref{ZETA}) shows that there is a distinguished numerical value of the coordinate $\psi$: causality puts an upper bound on it, given by
\begin{equation}\label{ETA}
\Psi = \sqrt{L/a}.
\end{equation}
Now $\psi$ is a spacelike coordinate, so of course neither it nor any of its specific values should be complexified when we pass to the Euclidean geometry. Thus $\Psi$ does not change. [As $v(x)$ is a time derivative, it complexifies, under passage to the Euclidean version, as $v(x) \rightarrow -i v(x)$, that is, in the same way as $a$; so both sides of equation (\ref{ZETA}) transform in the same way, since, again, $\psi$ does not change.]

The Euclidean norm of $\partial_t$ is computed with respect to the Euclidean QKMV$_0$, or ``EQKMV$_0$'' metric:
\begin{equation}\label{IOTA}
g(EQKMV_0)(\partial_t,\partial_t)\;=\;{- \Delta^E_r \;-\;a^2\Delta^E_{\psi} \over \rho_E^2},
\end{equation}
where
\begin{eqnarray}\label{KAPPA}
\rho_E^2& = & r^2\;-\;a^2\psi^2 \nonumber\\
\Delta^E_r & = & -a^2+ {r^4\over L^2} - 8\pi M^* r - 4\pi Q^{*2}\nonumber\\
\Delta^E_{\psi}& = & 1 - {a^2 \psi^4\over L^2}.
\end{eqnarray}
Notice, from the last of these equations, that $\Delta^E_{\psi}$, unlike its Lorentzian counterpart, \emph{can} vanish; and that $\Psi$, defined earlier by equation (\ref{ETA}), is also the maximal value of $\psi$ in the Euclidean case. [Notice too that, from the second of these equations, $r$ always satisfies $r^4/L^2 > a^2$ for all $r \geq r^E_h$, where the latter is defined by $\Delta^E_r(r^E_h) = 0$; we only consider such values of $r$. But this means that $r^2 > aL$, or in other words $r^2 > a^2\Psi^2$, which means that $\rho_E$ never vanishes.]

The right side of equation (\ref{IOTA}) vanishes precisely when $r = r^E_h$, and at the same time $\psi = \Psi$. Hence, in the Euclidean case, $\partial_t$ vanishes at those values of $r$ and $\psi$ [because, in Euclidean geometry, a vector must vanish if its norm does so]. One can easily see that the same holds true of $\partial_{\zeta}$.

The Euclidean version of the electromagnetic potential is
\begin{equation}\label{LAMBDA}
A^E = \Bigg[-{Q^*r \over  \rho_{E}^2}+\kappa_t^E\Bigg]\m{d} t \;+\; \Bigg[{aQ^*r\psi^2 \over  \rho_{E}^2}+\kappa_{\zeta}^E\Bigg]\m{d}\zeta,
\end{equation}
where $\kappa_t^E$ and $\kappa_{\zeta}^E$ are constants. We see that, since we showed that $\rho_E$ never vanishes, $A^E$ is regular everywhere, for any values of these constants, \emph{except} possibly at those points where $\partial_t$ and $\partial_{\zeta}$ vanish. That happens at $r = r^E_h, \psi = \Psi$, so that $A^E(\partial_t) =  A^E(\partial_{\zeta}) = 0$ there if $A^E$ is to be regular everywhere. This will allow us to fix the constants $\kappa_t^E$ and $\kappa_{\zeta}^E$, and they can then be continued to obtain the corresponding constants in the Lorentzian case. Thus we can fix the asymptotic potential by requiring the regularity of \emph{both} the Lorentzian and the Euclidean potential forms.

Carrying out the continuation [replacing  $t \rightarrow it$,  $a \rightarrow -ia$,  $Q^* \rightarrow -iQ^*$], we obtain
\begin{equation}\label{MU}
A = \Bigg[-{Q^*r \over  \rho^2}\;+\;{Q^*r_h\over r_h^2+a^2\Psi^2}\Bigg]\m{d} t \;+\; \Bigg[-{aQ^*r\psi^2 \over  \rho^2}+{aQ^*r_h\Psi^2\over r_h^2+a^2\Psi^2} \Bigg]\m{d}\zeta.
\end{equation}
Letting $r \rightarrow \infty$ and using equation (\ref{ETA}), we now obtain the asymptotic electromagnetic potential:
\begin{equation}\label{NU}
A_{\infty} = {Q^*r_h\over r_h^2+aL}[\m{d}t \;+\; \m{d}z],
\end{equation}
having set $z = L\zeta$. Finally, $\mu$ is obtained \cite{kn:koba} as a multiple\footnote{See \cite{kn:myers}; we follow the same conventions.}  of the asymptotic value of the time component of $A$. We have
\begin{equation}\label{OMICRON}
\mu \;=\;{Q^*r_h \over \pi L [r_h^2 + aL]}.
\end{equation}

Given the physical parameters $L, T, a$, and $\mu$ in the boundary field theory, we can solve the three equations, (\ref{EPSILON2}),(\ref{ZETA2}), and (\ref{OMICRON}), for the three geometric parameters $M^*$, $Q^*$, and $r_h$, and so we can completely specify the dual QKMV$_0$ bulk geometry.

Representative values are as follows. For any asymptotically AdS spacetime, the curvature parameter $L$ has a physical interpretation as a \emph{time} scale: the proper time required for an object to fall from infinity to any fixed location in AdS itself is given \cite{kn:gibbo} by $\pi$L/2, and similar formulae apply in more general asymptotically AdS cases. We see that $L$ should therefore be associated with some characteristic time scale of the QGP. A natural choice in this context is the lifetime of the QGP in the aftermath of a typical heavy-ion collision, roughly 5 femtometres in natural units.

The temperature at the critical point is thought to be around 175 MeV, or rather below 1 fm$^{- 1}$ in natural units. The temperature of the QGP might take values, in future beam energy scan experiments, very roughly\footnote{The beam scan experiments probe lower temperatures, but note that we are only interested in collisions which definitely produce a plasma, since of course only these could have a dual description in terms of a black hole.} in the range 0.5 - 2 fm$^{- 1}$. [Bear in mind that the QGP may be able to exist at relatively low temperatures when the chemical potential is large.]

In \cite{kn:75} we computed a rough estimate for the parameter $a$, in collisions with optimal impact parameter, attained in previous runs at the RHIC experiment: it is of the order 20 fm. It is not yet clear what values the beam energy scan programmes mentioned in Section 1 will produce, but, in order to be definite, we shall continue to use this value. [Actually, our results are remarkably insensitive to the value of this parameter when the chemical potential is not negligible; most of the numbers we obtain in that case would change by less than 1$\%$ if $a$ = 100 fm were used instead of $a$ = 20 fm.]

Typical values for $\mu$ around the critical point are estimated to be similar to the temperature; one can hope that the experiments will ultimately probe values for $\mu$ ranging between 1 and perhaps 3 or 4 fm$^{- 1}$. All of these estimates are provided mainly as an aid to intuition; we make no pretence of either accuracy or precision, which would be pointless under the circumstances. The point is that, given such data, we can assign definite numerical values to all of the geometric bulk parameters.
Later we will see how to use this correspondence.

Now we turn to the problem of finding the dual of the shear-induced instabilities which may exist in the QGP.

\addtocounter{section}{1}
\section* {\large{\textsf{4. The Dual of the Shearing Instability}}}
Our objective now is to find a dual version of the shearing instability which is conjectured \cite{kn:KelvinHelm} to arise in the QGP generated by a peripheral heavy-ion collision. Specifically, we need to find a form of AdS black hole instability which is [a] triggered by the presence of angular momentum arising in connection with shear, but [b] \emph{not} triggered by purely rotational angular momentum, since fluid rotation does not give rise to instabilities of the Kelvin-Helmholtz type.

We claim that the form of instability studied by Seiberg and Witten \cite{kn:seiberg}, in quite general asymptotically AdS spacetimes, meets these criteria. This mechanism is based on the study of a certain kind of BPS branes wrapping around submanifolds in such spacetimes. The action consists of both positive and negative terms, which compete as one considers increasingly remote branes. In AdS itself, and in the AdS-Schwarzschild geometry, the positive term [related to the area of the brane] dominates, and so the system is stable; but if the geometry is distorted [by the addition of charge or angular momentum, or by changing the topology of the event horizon] then the negative term [computed from the brane volume] can cause the action to become unbounded below, and the system becomes unstable. Seiberg and Witten show that the instability can be explained dually in terms of the conformal coupling of a certain scalar field. [A similar phenomenon has been explored in detail in \cite{kn:yamada1}\cite{kn:yamada2}.]

Our plan is to set up such a system on the QKMV$_0$ background. We do not claim that this system \emph{necessarily} arises in such a background: we are artificially constructing it [in the ``bottom up'' manner] so as to find a dual description of the shearing instability in the QGP. Similarly, we do not claim that QCD itself has a conformal scalar: we are merely using the Seiberg-Witten scalar as a well-understood toy model to represent the triggering of hydrodynamic instability in the QGP. The hope is that this will lead to less crude models of this triggering when more realistic holographic models of QCD become available.

Let us turn, then, to the form taken by the Seiberg-Witten brane action when the background is given by the QKMV$_0$ geometry. Essentially this involves a somewhat intricate computation of the volume and area of a brane inhabiting the Euclidean version of the QKMV$_0$ geometry; the brane wraps around $r$ = constant sections, and so the brane action is a function of $r$. It is a constant multiple of the difference between the area of such a section and its volume [with the latter corrected by a factor of $3/L$, see \cite{kn:seiberg}]. One then continues back to Lorentzian signature.

We begin with the simplest case, in which both the angular momentum and the electric charge are zero, so that we are dealing with the Planar AdS-Schwarzschild geometry [obtained by setting $Q^*$ = 0 in equation (\ref{ALPHA})]. We have, up to a constant factor which need not detain us, a brane action of the form
\begin{equation}\label{eq:PI}
\$(PAdSS) \;\sim \; \Bigg[{r^2\over L^2}-{8\pi M^*\over r}\Bigg]^{1/2}r^2\;-\;{3\over L}\int_{r_h}^rr^2\,dr,
\end{equation}
where the first term arises from the area of the brane [which of course extends also in the $t$ direction], and the second is related to the volume [which has a simpler form since the $tt$ and $rr$ components of the metric are related reciprocally in this case]. This can be simplified to
\begin{equation}\label{eq:RHO}
\$(PAdSS) \;\sim \; {r^3_h\over L}\;-\;{8\pi M^*L\over 1+\Big[1-{8\pi M^*L^2\over r^3}\Big]^{1/2}}.
\end{equation}
It is easy to see that this is a function which vanishes at the event horizon, but increases monotonically from there: a brane nucleating in this geometry will tend to contract back into the black hole; so the system is \emph{stable} in the absence of angular momentum.

If we repeat this calculation\footnote{See \cite{kn:75} for the details of such calculations. In order to obtain a finite result, one needs to perform the computation on a compact Euclidean manifold. The compactification of the Euclidean ``time'' coordinate is needed in any case, for the usual reasons, and this actually forces the Euclidean version of $\zeta$ to be compact when $a$ does not vanish. We saw earlier that the Euclidean version of $\psi$ is also compact.} for the full QKMV$_0$ geometry, with neither charge nor angular momentum equal to zero, then we find, again disregarding a constant multiplicative factor, that the action is given by
\begin{eqnarray}\label{SIGMA}
\$(QKMV_0) \; & \sim & \;{r^2\over 2} \sqrt{a^2+ {r^4\over L^2} - 8\pi M^*r+4\pi Q^{*2}}\Bigg[{1 \over a}\m{arcsinh}\Big({\sqrt{aL}\over r}\Big) + {\sqrt{L/a} \over r}\sqrt{1+{aL \over r^2}}\Bigg] \nonumber \\
& & \;\;\;\;\;\;\;\;\;\;\;\;\;\;\;\;\;\;\;\; -  {1\over L}\Bigg[\sqrt{{L \over a}}(r^3 - r_{h}^3) + \sqrt{aL^3}(r-r_{h})\Bigg].
\end{eqnarray}
As before, the action vanishes at the event horizon, and it begins to increase. But eventually the shape changes, as can be seen by writing the action in the following form:
\begin{eqnarray}\label{eq:TAU}
\$(QKMV_0) \;&\sim&\; { -\,5\sqrt{aL} \over 6}\,r \;+\;\Bigg[\,{r_{h}^3 \over \sqrt{aL}} \;+\;\sqrt{aL}r_{h} \;-\;4\pi M^*\sqrt{L^3/a}\Bigg] \nonumber \\
& & \;\;\;\;\;\;\;\;\;\;\;\;\;\;\;\;\;\;\;\; \;+\; {(a^2 +4\pi Q^{*2})\sqrt{L^3/a}\over 2r}\;+\;O(1/r^2) .
\end{eqnarray}
One sees from this equation that the action rather quickly comes to resemble a linear function with a negative slope proportional to the square root of the parameter $a$. The shape of a typical graph is given in Figure 2.

\begin{figure}[!h]
\centering
\includegraphics[width=0.8\textwidth]{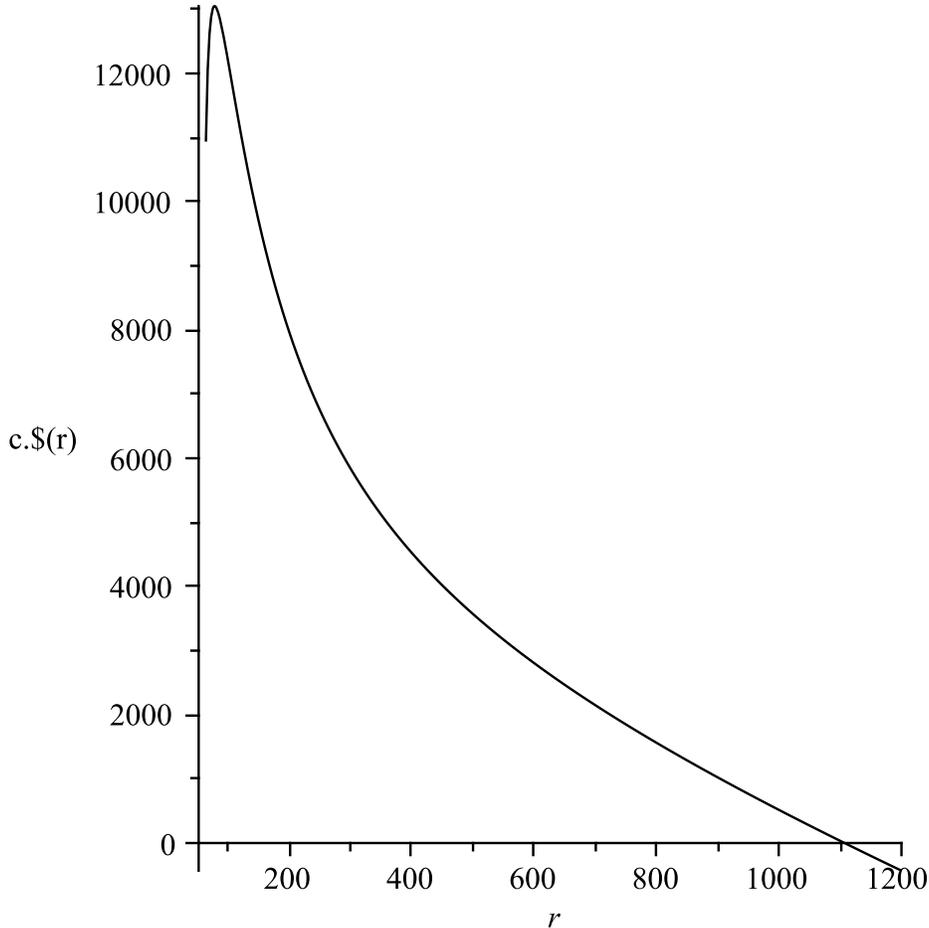}
\caption{[Multiple of the] Brane Action,T = 2 fm$^{-1}$,  $\mu$ = 1 fm$^{-1}$, a = 100 fm, L = 5 fm.}
\end{figure}

The conclusion is that, for all non-zero angular momenta, the action will become and remain \emph{negative} beyond some value of $r = r_{neg}$; it is in fact unbounded below. This is not due to the presence of electric charge: \emph{it is caused by the presence of the parameter} $a$. We see that, as claimed, a Seiberg-Witten brane system can detect shearing angular momentum. It does so by setting off an instability in the bulk: a brane which nucleates in a region \emph{sufficiently far away} from the black hole will always be able to decrease its action globally [below the value near the horizon] by expanding; it can escape to infinity. This corresponds to a dual instability in the field theory at infinity. If we interpret the field theory as representing the QGP fluid, this instability is to be interpreted as representing a typical hydrodynamic instability generated by fluid shear, similar to the Kelvin-Helmholtz instability \cite{kn:KelvinHelm}.

The scale on the vertical axis in Figure 2 has no significance, but the horizontal scale does. We see that $r_{neg}$ is very large [relative to the length scales discussed above], and this is typical. This aspect of the instability is useful, because it shows that the effect is robust. For example, we cannot hope to show that couplings other than that of the brane to the metric [such as might arise, in the charged case, if one were able to obtain the electric field by means of a Kaluza-Klein reduction of some higher-dimensional theory] will remove the instability. To do so, they would have to be effective in the asymptotic region, where there should be no ``hair'' other than the usual ADM parameters of the black hole metric; this is highly implausible\footnote{Furthermore, they would have to remove an infinite quantity of negative action.}. We conclude that, certainly at angular momenta and charges which are not extremely large, we can analyse the instability just by studying the effect of the spacetime geometry on the structure of the pure brane action. [This was the procedure followed, for example, by Maldacena and Maoz \cite{kn:maoz}, who were the first to consider Seiberg-Witten instability in the presence of a gauge field.]

We see, then, that the Seiberg-Witten system does provide a concrete way of representing the triggering of instability by shear: the non-shearing planar AdS-Schwarzschild black hole is stable, but it becomes unstable as soon as shearing, represented by the QKMV$_0$ deformation of the geometry, is introduced. The dual statement is that the fluid at infinity becomes unstable under shear.

But what happens when the fluid angular momentum is due to \emph{rotation} rather than shear? This corresponds, in the bulk, to the topologically spherical AdS-Kerr-Newman geometry. To see this, note first that, at infinity, the AdS-Kerr-Newman geometry gives rise to a rigidly rotating space described [see for example \cite{kn:sonner}] by the metric
\begin{equation}\label{TAUTAU}
g(AdSKN)_{\infty}\;=\;-\,\m{d}t^2 \;-\;{2a\,\m{sin}^2(\theta)\,\m{d}t \m{d}\phi\over \Xi} \;+\; {L^2 \, \m{d}\theta^2 \over 1 - (a/L)^2\m{cos}^2(\theta)} \;+\; {L^2 \m{sin}^2(\theta)\m{d}\phi^2\over \Xi},
\end{equation}
where $\theta$ and $\phi$ are the usual spherical coordinates, where $L$ and $a$ have their usual meanings, and where $\Xi = 1 - (a^2/L^2)$, so that $a$ must be strictly less than $L$ in this geometry. [One would therefore need to revise our value of $L$ if one wished to explore this further; this will not affect our argument here, however.] If we are near to the pole at $\theta = 0$, so that cos($\theta$) can be approximated by unity, then the last two terms in this metric can be interpreted as polar coordinates on a round sphere of radius $L/\sqrt{\Xi}$; if this number is sufficiently large, then one can approximate the sphere near the pole by its tangent plane there, so that, if we define $\varrho = L\theta/\sqrt{\Xi}$, then ($\varrho, \phi$) can be regarded as polar coordinates in the plane. Because of the cross-term involving d$t$ and d$\phi$, this plane is rotating rigidly, at a rate controlled by $a$. Thus we obtain a space in which the rigidly rotating planar fluid can be modelled.

However, the AdS-Kerr-Newman black hole is immune to Seiberg-Witten instability. This can be seen most directly by computing the scalar curvature of the [Euclidean version of] the boundary metric: it is given by
\begin{equation}\label{TAUTAUTAU}
R(g(EAdSKN_{\infty})) \;=\; 2\Big(1\;-\;{a^2 \over L^2}\Big)\;+\;10\,{a^2 \over L^2}\,\m{cos}^2(\theta),
\end{equation}
which is strictly positive, since $a$ must be smaller than $L$. This means that the conformally coupled scalar on the boundary effectively receives a positive contribution to its squared mass, so it remains stable. As Seiberg and Witten explain in detail, this condition also implies the stability of the bulk, in the sense that the brane action is at least asymptotically positive. Thus our model is able to distinguish between shearing and rotational angular momentum: it predicts that the former can give rise to instabilities in the boundary fluid, but that the latter cannot. This is precisely in accord with fluid mechanical expectations.

Even in the shearing case, however, it is not clear that the instability is physically relevant: all instabilities take time to evolve, and this time may not be available ---$\,$ a point which often plays a crucial role in heavy ion physics [see \cite{kn:ippy} for an example]. We now propose a very simple way of roughly estimating the time required for the Seiberg-Witten instability to make its presence felt.

\addtocounter{section}{1}
\section* {\large{\textsf{5. Evolution of SW Instability}}}
It is clear that the time required for the Seiberg-Witten instability to manifest itself in these systems should be related to the time required for the region of negative action [which begins at $r = r_{neg}$] to influence the entire bulk, including the region near to [but outside] the event horizon. We compute this as the proper time required for a massive but electrically neutral particle, released from rest relative to a zero angular momentum observer, to fall from a point with $r = r_{neg}$ to $r = r_h$, the location of the event horizon.

For simplicity we consider the case of orbits in the QKMV$_0$ geometry satisfying  $\psi$ = 0, the ``equatorial" plane. Then the equations for the orbit are, as in \cite{kn:75},
\begin{equation}\label{eq:UPSILON}
\dot{t}\Bigg[{-\,\Delta_r \over r^2}\;+\;{a^2 \over r^2}\Bigg]\;-\;\dot{\zeta}a\;=\;c,
\end{equation}
where a dot denotes a derivative with respect to proper time, and where $c$ is a constant, and
\begin{equation}\label{eq:PHI}
-\,\dot{t}a\;+\;\dot{\zeta}r^2\;=\;0.
\end{equation}
Combining these with the fact that the geodesic tangent is of unit length and with the initial conditions, one finds that the time required, which we denote by $\tau(a, T, \mu, L)$, is given by
\begin{equation}\label{XI}
\tau(a, T, \mu, L)\;=\;\int_{r_{h}}^{r_{neg}} \Bigg[{a^2 + 4\pi Q^{*2}\over r_{neg}^2}\,+\,{r_{neg}^2 \over L^2}\,-\,{8\pi M^*\over
r_{neg}}\,-\,{a^2 + 4\pi Q^{*2}\over r^2}\,-\,{r^2 \over L^2}\,+\,{8\pi M^*\over
r}\Bigg]^{-\,1/2}dr.
\end{equation}
Given the field theory parameters $a, T, L,$ and $\mu$, we can compute, as we have discussed, $r_h$, $M^*$, and $Q^*$, so $r_{neg}$ can be computed by finding the larger positive real root of the expression in equation (\ref{SIGMA}). Thus everything in this expression is now known to us in terms of the boundary data.

We can now numerically investigate $\tau$(a, T, $\mu$, L) as a function of position in the quark matter phase diagram, specifically in the region 0.5 $<$ T $<$ 2, 0 $<$ $\mu$ $<$ 5 in units of inverse fm. We use the parameter values discussed earlier, that is, $a = 20$ fm, $L = 5$ fm. The results are shown in the table: the times are given as multiples of the approximate lifetime of the QGP in heavy ion collisions, so that values below unity mean that the instability can develop within the normal lifetime of the plasma\footnote{We remind the reader that the T = 1/2, $\mu$ = 0 entry is not physical, because the plasma does not exist in that corner of the phase diagram. The same may be true of the T = 1/2, $\mu$ = 1 entry.}.

\begin{center}
\begin{tabular}{|c|c|c||c|c||c|c|}
  \hline
 & $\mu$ = 0  &  $\mu$ = 1 & $\mu$ = 2 &$\mu$ = 3 &$\mu$ = 4&$\mu$ = 5 \\
\hline
T = 2 &    1.5727    &  1.5621 & 1.1219 &0.9164&0.8183  &0.7623\\
T = 1 &   1.5615  &  1.1178 & 0.8178&0.7263&0.6834 &0.6588\\
T = 1/2 &  1.5195  &  0.8161 & 0.6832&0.6429&0.6238&0.6129 \\
\hline
\end{tabular}
\end{center}

We do not feel that the numerical results themselves should be taken very seriously; instead, let us focus on the \emph{trends} in the results. The principal observation from that point of view is that, as one scans across the table to larger values of the chemical potential, the instability develops at a dramatically enhanced rate, effectively shortening the lifetime of the boundary field theory as a stable hydrodynamic system. If these results can be extrapolated to the fluid behaviour of the QGP, then this means that the scan to larger values of $\mu$ will probe plasmas with a shortened lifetime as a stable fluid undergoing laminar flow.

Quark polarization is thought \cite{kn:huang} to be a cumulative process, building up as a result of multiple scattering; therefore it requires a substantial period of time, relative to the lifetime of the plasma, to reach observable levels, if it does so. It is possible that the early onset of hydrodynamic instability interferes with this process, explaining the non-observation of quark polarization thus far. If that is so, then the message from holography is that plasmas at high values of $\mu$ are even less likely to give rise to polarization than those which have already been investigated.

\addtocounter{section}{1}
\section* {\large{\textsf{6. Conclusion: Quark Polarization in the Phase Diagram}}}
Peripheral heavy-ion collisions give rise to a shearing quark plasma with very large angular momentum densities; this should lead to distinctive observational signatures involving quark polarization. But fluid shear often leads to hydrodynamic instability, which might disrupt such effects. In short, the angular momentum may conceal its own presence. That will be the case if the instability develops quickly. 

In this work we have attempted to obtain a holographic model of the fluid shear in the QGP, and of the instability it induces. This is done by deliberately constructing a system of stable branes in a planar AdS-Schwarzschild black hole spacetime, and seeing what happens when the bulk is endowed with either shearing or rotational angular momentum. We find that the system remains stable in the case of rotation, but that it develops an instability under shearing. This is in agreement with the behaviour of a fluid at infinity, giving us some confidence that this is the right way to probe the holography of dual fluids with angular momentum.

Next, we suggested a holographic method of estimating the time required for the instability to develop in the shearing case. By modifying the charge on the black hole, we can study the effects of varying the quark chemical potential. We find that holography predicts that instability will set in earlier at large values of $\mu$. Although this is only a qualitative prediction, it may nevertheless be falsifiable: we do \emph{not} expect quark polarization effects to become more apparent in the beam energy scan experiments. There is however still reason to hope that they will be found in the region of the quark matter phase diagram being probed by the ALICE experiment \cite{kn:reygers}, that is, at high temperatures but small values of $\mu$.

\addtocounter{section}{1}
\section*{\large{\textsf{Acknowledgements}}}
The author is very grateful for the kind hospitality of NORDITA, for providing facilities and an environment in which this work could be done, and wishes to thank particularly L$\acute{\m{a}}$rus Thorlacius and the organisers of the ``Holographic Way'' programme. He also wishes to thank Soon Wanmei for support, encouragement, and advice on numerical questions.


\begin{thebibliography}{18}
\bibitem{kn:behan}
Sourendu Gupta, Xiaofeng Luo, Bedangadas Mohanty, Hans Georg Ritter, Nu Xu, Scale for the Phase Diagram of Quantum Chromodynamics,
Science 332 (2011) 1525, arXiv:1105.3934 [hep-ph]
\bibitem{kn:ohnishi}
Akira Ohnishi, Phase diagram and heavy-ion collisions: Overview, Prog.Theor.Phys.Suppl. 193 (2012) 1, arXiv:1112.3210 [nucl-th]
\bibitem{kn:enrod}
G. Endrodi, QCD phase diagram: overview of recent lattice results, arXiv:1311.0648 [hep-lat]
\bibitem{kn:ilya}
Ilya Selyuzhenkov, Recent experimental results from the relativistic heavy-ion collisions at LHC and RHIC, arXiv:1109.1654 [nucl-ex]
\bibitem{kn:dong}
Xin Dong (for the STAR Collaboration), Highlights from STAR, Nucl.Phys.A904-905 2013 (2013) 19c, arXiv:1210.6677 [nucl-ex]
\bibitem{kn:fair}
M. Bleicher, M. Nahrgang, J. Steinheimer, P. Bicudo, Physics Prospects at FAIR, Acta Phys.Polon. B43 (2012) 731, arXiv:1112.5286 [hep-ph]
\bibitem{kn:nica}
V. D. Kekelidze, A. D. Kovalenko, I. N. Meshkov, A. S. Sorin, G. V. Trubnikov, NICA at JINR: New prospects for exploration of quark-gluon matter,
Physics of Atomic Nuclei 75 (2012) 542
\bibitem{kn:fuku}
Kenji Fukushima, Chihiro Sasaki, The phase diagram of nuclear and quark matter at high baryon density, Prog.Part.Nucl.Phys. 72 (2013) 99, arXiv:1301.6377 [hep-ph]
\bibitem{kn:katz}
Zoltan Fodor, Sandor D. Katz, Christian Schmidt,
The Density of states method at non-zero chemical potential, JHEP 0703:121,2007, \x hep-lat/0701022
\bibitem{kn:enrodi}
G. Endrodi, Z. Fodor, S.D. Katz, K.K. Szabo,
The QCD phase diagram at nonzero quark density,
JHEP 1104 (2011) 001, arXiv:1102.1356 [hep-lat]
\bibitem{kn:andronic}
A. Andronic, D. Blaschke, P. Braun-Munzinger, J. Cleymans, K. Fukushima, L.D. McLerran, H. Oeschler, R.D. Pisarski, K. Redlich, C. Sasaki, H. Satz, J. Stachel,
Hadron Production in Ultra-relativistic Nuclear Collisions: Quarkyonic Matter and a Triple Point in the Phase Diagram of QCD, Nucl.Phys. A837 (2010) 65, 0911.4806 [hep-ph]
\bibitem{kn:borun}
Jan de Boer, Borun D. Chowdhury, Michal P. Heller, Jakub Jankowski, A holographic realization of the Quarkyonic phase,  Phys. Rev. D 87, 066009 (2013), arXiv:1209.5915 [hep-th]
\bibitem{kn:liang}
Zuo-Tang Liang, Xin-Nian Wang, Globally polarized quark-gluon plasma in non-central A+A collisions,
Phys.Rev.Lett. 94 (2005) 102301, Erratum-ibid. 96 (2006) 039901, \x nucl-th/0410079
\bibitem{kn:bec}
F. Becattini, F. Piccinini, J. Rizzo, Angular momentum conservation in heavy ion collisions at very high energy, Phys.Rev.C77:024906,2008,
arXiv:0711.1253 [nucl-th]
\bibitem{kn:huang}
Xu-Guang Huang, Pasi Huovinen, Xin-Nian Wang, Quark Polarization in a Viscous Quark-Gluon Plasma,
Phys. Rev. C84, 054910(2011), arXiv:1108.5649 [nucl-th]
\bibitem{kn:bemo}
Victor Roy, A. K. Chaudhuri, Bedangadas Mohanty,
Comparison of results from a 2+1D relativistic viscous hydrodynamic model to elliptic and hexadecapole flow of charged hadrons measured in Au-Au collisions at $\sqrt{s_{\rm {NN}}}$ = 200 GeV, Phys.Rev. C86 (2012) 014902, arXiv:1204.2347 [nucl-th]
\bibitem{kn:leigh}
Marco M. Caldarelli, Robert G. Leigh, Anastasios C. Petkou, P.Marios Petropoulos, Valentina Pozzoli, Konstadinos Siampos, 	
Vorticity in holographic fluids, PoS CORFU2011 (2011) 076, arXiv:1206.4351
\bibitem{kn:sorin}
Mircea Baznat, Konstantin Gudima, Alexander Sorin, Oleg Teryaev, Helicity separation in Heavy-Ion Collisions, arXiv:1301.7003 [nucl-th]
\bibitem{kn:STAR}
J H Chen (for the STAR Collaboration),
Spin alignment of K$^{*0}$(892) and $\phi$ (1020) mesons in Au+Au and p+p collisions at $\sqrt{s_{\rm NN}}$ = 200 GeV,
J. Phys. G: Nucl. Part. Phys. 35 (2008) 044068
\bibitem{kn:RHIC}
STAR Collaboration (B.I. Abelev et al.),
Spin alignment measurements of the K$^{*0}$(892) and $\phi$ (1020) vector mesons in heavy ion collisions at $\sqrt{ s_{NN}}$ = 200 GeV,
Phys.Rev. C77 (2008) 061902, arXiv:0801.1729 [nucl-ex]
\bibitem{kn:jianhua}
Jian-Hua Gao, Shou-Wan Chen, Wei-tian Deng, Zuo-Tang Liang, Qun Wang, Xin-Nian Wang,
Global quark polarization in non-central A+A collisions, Phys.Rev. C77 (2008) 044902, arXiv:0710.2943 [nucl-th]
\bibitem{kn:ayala}
Alejandro Ayala, Eleazar Cuautle, G. Herrera Corral, J. Magnin, Luis Manuel Montano, Spin alignment of vector mesons in heavy ion and proton - proton collisions, Phys.Lett. B682 (2010) 408, arXiv:0906.4295 [hep-ph]
\bibitem{kn:drazin}
P.G.Drazin and W.H.Reid, \emph{Hydrodynamic Stability}, Second Edition, Cambridge University Press, 2004.
\bibitem{kn:lin}
C.-C. Lin, \emph{The Theory of Hydrodynamic Stability}, Cambridge University Press, 1966.
\bibitem{kn:KelvinHelm}
L.P.Csernai, D.D.Strottman and Cs.Anderlik, Kelvin-Helmholz instability in high energy heavy ion collisions, Phys. Rev. C 85 (2012) 054901, arXiv:1112.4287 [nucl-th]
\bibitem{kn:solana}
Jorge Casalderrey-Solana, Hong Liu, David Mateos, Krishna Rajagopal, Urs Achim Wiedemann,
Gauge/String Duality, Hot QCD and Heavy Ion Collisions, arXiv:1101.0618 [hep-th]
\bibitem{kn:pedraza}
Mariano Chernicoff, J. Antonio Garcia, Alberto Guijosa, Juan F. Pedraza,
Holographic Lessons for Quark Dynamics, J.Phys.G G39 (2012) 054002, arXiv:1111.0872 [hep-th]
\bibitem{kn:youngman}
Youngman Kim, Ik Jae Shin, Takuya Tsukioka, Holographic QCD: Past, Present, and Future, Prog.Part.Nucl.Phys. 68 (2013) 55, arXiv:1205.4852 [hep-ph]
\bibitem{kn:carter}
B. Carter, Hamilton-Jacobi and Schrodinger separable solutions of Einstein's equations,
Commun.Math.Phys. 10 (1968) 280
\bibitem{kn:hawrot}
S.W. Hawking, C.J. Hunter, Marika Taylor, Rotation and the AdS/CFT correspondence, Phys.Rev.D59:064005,1999, \x hep-th/9811056
\bibitem{kn:schalm}
A. Nata Atmaja, K. Schalm, Anisotropic Drag Force from 4D Kerr-AdS Black Holes, JHEP 1104:070,2011,
arXiv:1012.3800 [hep-th]
\bibitem{kn:klemm}
D. Klemm, V. Moretti, L. Vanzo, Rotating Topological Black Holes, Phys.Rev.D57:6127,1998; Erratum-ibid.D60:109902,1999,
arXiv:gr-qc/9710123
\bibitem{kn:seiberg}
Nathan Seiberg, Edward Witten, The D1/D5 System And Singular CFT,
JHEP 9904 (1999) 017, \x hep-th/9903224
\bibitem{kn:75}
Brett McInnes, Fragile Black Holes and an Angular Momentum Cutoff in Peripheral Heavy Ion Collisions, Nucl. Phys. B861 (2012) 236, arXiv:1201.6443 [hep-th]
\bibitem{kn:76}
Brett McInnes, Universality of the Holographic Angular Momentum Cutoff, Nucl.Phys. B864 (2012) 722,  arXiv:1206.0120 [hep-th]
\bibitem{kn:ipp}
Andreas Ipp, Antonino Di Piazza, Jorg Evers, Christoph H. Keitel,
Photon polarization as a probe for quark-gluon plasma dynamics, Phys.Lett. B666 (2008) 315, arXiv:0710.5700 [hep-ph]
\bibitem{kn:reygers}
ALICE Collaboration (Klaus Reygers for the collaboration), A Quick Tour of Ultra-Relativistic Heavy-Ion Physics at the LHC,
arXiv:1208.1626 [nucl-ex]
\bibitem{kn:mateos}
David Mateos,
Gauge/string duality applied to heavy ion collisions: Limitations, insights and prospects, J.Phys.G G38 (2011) 124030, arXiv:1106.3295 [hep-th]
\bibitem{kn:karch}
Andreas Karch,
Recent Progress in Applying Gauge/Gravity Duality to Quark-Gluon Plasma Physics,  AIP Conf.Proc. 1441 (2012) 95, arXiv:1108.4014 [hep-ph]
\bibitem{kn:kovch}
Yuri V. Kovchegov, CFT applications to relativistic heavy ion collisions: a brief review, Rept.Prog.Phys. 75 (2012) 124301, arXiv:1112.5403 [hep-ph]
\bibitem{kn:shipu}
Jian-Hua Gao, Zuo-Tang Liang, Shi Pu, Qun Wang, Xin-Nian Wang, 	
Chiral Anomaly and Local Polarization Effect from Quantum Kinetic Approach, Phys.Rev.Lett. 109 (2012) 232301, arXiv:1203.0725 [hep-ph]
\bibitem{kn:chamblin}
Andrew Chamblin, Roberto Emparan, Clifford V. Johnson, Robert C. Myers, Thermodynamics and fluctuations of charged AdS black holes, Phys.Rev. D60 (1999) 104026,
hep-th/9904197
\bibitem{kn:koba}
Shinpei Kobayashi, David Mateos, Shunji Matsuura , Robert C. Myers , Rowan M. Thomson, Phase transitions at finite baryon density,
JHEP 0702 (2007) 016, hep-th/0611099 [hep-th]
\bibitem{kn:lemmo}
J.P.S. Lemos, Phys.Lett.B353:46,1995,
\x gr-qc/9404041; R.B. Mann, Class.Quant.Grav. 14 (1997) L109, arXiv:gr-qc/9607071;
Rong-Gen Cai, Yuan-Zhong Zhang, Phys.Rev.D54:4891,1996, \x gr-qc/9609065;
Danny Birmingham, Class.Quant.Grav. 16 (1999) 1197, arXiv:hep-th/9808032
\bibitem{kn:peldan}
Dieter R. Brill, Jorma Louko, Peter Peldan,
Thermodynamics of (3+1)-dimensional black holes with toroidal or higher genus horizons,
Phys.Rev. D56 (1997) 3600, arXiv:gr-qc/9705012
\bibitem{kn:confined}
Edward Witten, Anti-de Sitter space, thermal phase transition, and confinement in gauge theories,
Adv.Theor.Math.Phys.2:505,1998, \x hep-th/9803131
\bibitem{kn:sonner}
Julian Sonner, A Rotating Holographic Superconductor, Phys.Rev.D80:084031,2009, arXiv:0903.0627 [hep-th]
\bibitem{kn:cognola}
Marco M. Caldarelli, Guido Cognola, Dietmar Klemm, Thermodynamics of Kerr-Newman-AdS Black Holes and Conformal Field Theories,
Class.Quant.Grav. 17 (2000) 399, arXiv:hep-th/9908022
\bibitem{kn:exact}
J.B. Griffiths and J. Podolsky, \emph{Exact Space-Times in Einstein's General Relativity}, Cambridge University Press, 2009.
\bibitem{kn:myers}
Robert C. Myers, Miguel F. Paulos, Aninda Sinha,
Holographic Hydrodynamics with a Chemical Potential, JHEP 0906:006,2009, arXiv:0903.2834
\bibitem{kn:gibbo}
G.W. Gibbons, Anti-de-Sitter spacetime and its uses, Proceedings of the 2nd Samos Meeting on Cosmology, Geometry and Relativity, S Cotsakis and G W Gibbons eds,  Lecture Notes in Physics 537 (2000), arXiv:1110.1206 [hep-th]
\bibitem{kn:yamada1}
Daiske Yamada, Metastability of R-charged black holes, Class.Quant.Grav.24:3347-3376,2007, \x hep-th/0701254
\bibitem{kn:yamada2}
Daiske Yamada, Fragmentation of Spinning Branes, Class.Quant.Grav.25:145006,2008,
arXiv:0802.3508 [hep-th]
\bibitem{kn:maoz}
Juan Maldacena, Liat Maoz, Wormholes in AdS, JHEP 0402 (2004) 053,
\x hep-th/0401024
\bibitem{kn:ippy}
Andreas Ipp, Unstable dynamics of Yang-Mills fields at early times of heavy ion collisions, J.Phys.Conf.Ser. 422 (2013) 012028, \x 1210.5150 [hep-ph]


\end{thebibliography}
\end{document}